\documentclass{Interspeech2024}
\ninept
\interspeechcameraready 
\usepackage[subtle]{savetrees} %subtle, moderate

\usepackage{setspace}
\usepackage{bibspacing}
\title{FoVNet: Configurable Field-of-View Speech Enhancement with Low Computation and Distortion for Smart Glasses}
\name[affiliation={1}]{Zhongweiyang}{Xu}
\name[affiliation={2}]{Ali}{Aroudi}
\name[affiliation={2}]{Ke}{Tan}
\name[affiliation={2}]{Ashutosh}{Pandey}
\name[affiliation={2}]{Jung-Suk}{Lee}
\name[affiliation={2}]{Buye}{Xu}
\name[affiliation={2}]{Francesco}{Nesta}

\address{
  $^1$University of Illinois Urbana-Champaign, IL, USA\\
  $^2$Reality Labs Research at Meta, WA, USA}
\email{zx21@illinois.edu, aliaroudi@meta.com, tanke1116@meta.com, apandey620@meta.com, jungsuklee@meta.com, xub@meta.com, francesconesta@meta.com}

\keywords{speech enhancement, array signal processing}

\begin{document}

\maketitle

\vspace{-50pt}
\begin{abstract}

\vspace{-3pt}
\begin{spacing}{0.95}
This paper presents a novel multi-channel speech enhancement approach, FoVNet, that enables highly efficient speech enhancement within a configurable field of view (FoV) of a smart-glasses user without needing specific target-talker(s) directions. It advances over prior works by enhancing all speakers within any given FoV, with a hybrid signal processing and deep learning approach designed with high computational efficiency. The neural network component is designed with ultra-low computation (about 50 MMACS). A multi-channel Wiener filter and a post-processing module are further used to improve perceptual quality. We evaluate our algorithm with a microphone array on smart glasses, providing a configurable, efficient solution for augmented hearing on energy-constrained devices. FoVNet excels in both computational efficiency and speech quality across multiple scenarios, making it a promising solution for smart glasses applications.
\end{spacing}
    % Manuscripts submitted to Interspeech 2024 must use this document as both an instruction set and as a template. Do not use a past paper as a template. Always start from a fresh copy, and read it all before replacing the content with your own. The main changes with respect to previous instructions are \red{highlighted in red}.
    
    % Before submitting, check that your manuscript conforms to this template. If it does not, it may be rejected. Do not be tempted to adjust the format! Instead, edit your content to fit the allowed space. The maximum number of manuscript pages is 5. The 5th page is reserved exclusively for \red{acknowledgements} and references, which may begin on an earlier page if there is space.
    
    % The abstract is limited to 1000 characters. The one in your manuscript and the one entered in the submission form must be identical. Avoid non-ASCII characters, symbols, maths, italics, etc as they may not display correctly in the abstract book. Do not use citations in the abstract: the abstract booklet will not include a bibliography.  Index terms appear immediately below the abstract. 
\end{abstract}

\vspace{-8pt}
\section{Introduction}
\vspace{-5pt}
% traditional methods
% Multi-channel speech enhancement aims to remove the noise while preserving the speech components given a microphone array recorded noisy signal. 

% Classical signal processing methods in this area are well-developed, including ICA~\cite{ICA}, NMF~\cite{NMF}, and beamforming~\cite{beamforming}. These classical methods are more or less sub-optimal due to the assumptions they make.

% multi-channel speech enhancement
\begin{spacing}{0.95}

Multi-channel speech enhancement aims at denoising the target speech given a microphone array recorded noisy signal. With the success of deep learning, neural multi-channel speech enhancement techniques yield a huge performance boost over non-neural speech enhancement techniques. Nevertheless, the current neural multi-channel speech enhancement techniques need significant improvement in terms of performance, configurability, and computational cost to be effectively utilized in smart glasses applications for enhancing daily speech communication. Daily acoustic scenarios of speech communication can encompass a wide range of situations, including those involving a single talker or multiple talkers. For single-talker speech enhancement scenarios, several methods have been proposed. Time-domain methods~\cite{luo2019fasnet, llrnn, Pandey2019TCNNTC} directly apply temporal 1D-CNN on framed multi-channel audio. One established approach first uses a neural network to estimate a speech spectral mask and then uses this mask for MVDR beamforming~\cite{consolidated, mask_mvdr, mask_mvdr2, mask_mvdr3}. Following such an approach, ADL-MVDR method has been proposed aiming at estimating the MVDR beamforming outputs in an end-to-end fashion~\cite{adl-mvdr}. Another established approach exploits DNN1-Wiener-DNN2 method~\cite{dnnwienerdnn1, dnnwienerdnn2} in a modular way, where DNN1 estimates the target signal, a multi-channel wiener filter is then built based on the estimated target signal, and finally, DNN2 further enhances the Wiener filtered signal. \cite{e2ewiener} further explores this approach by training the whole system in an end-to-end fashion.

% region based separation
Acoustic scenarios can also involve multiple talkers, such as cocktail party scenarios, which pose challenges for speech enhancement, automatic speech recognition (ASR), and speaker diarization. In these complex acoustic scenes, distinguishing target talkers from interfering talkers for processing or enhancement is a non-trivial task. One common approach to address this challenge is direction-aware speech enhancement, where the direction of arrival (DoA) of the target talker is assumed to be known and used to compute DoA-dependent spatial features for extracting the target speech~\cite{spatial_feature1, spatial_feature2, Tesch_2024}. Direction-aware speech enhancement is effective for applications where the spatial region of enhancement is pre-defined, such as in-car scenarios or head-worn microphone array scenarios where the target directions of arrival (DoAs) are fixed \cite{binaural, zoneformer, DSENET, incar_new}. However, for scenarios where the spatial region of interest is flexible, techniques such as Cone of Silence (CoS) and ReZero have been recently proposed \cite{jenrungrot2020cone, gu2023rezero}. It is important to note that developing a neural network approach that is well-generalized for any arbitrary spatial region requires careful encoding of spatial information within the network, which can be computationally expensive. CoS~\cite{jenrungrot2020cone} first beamforms towards a single DoA and then uses one-hot vectors to represent a spatial width around that beamformed DoA. To encode any spatial region, Rezero~\cite{gu2023rezero} first samples different directions inside the region, and further designs a distance feature that allows the spatial region to be flexible along the distance axis. Although these approaches allow for region-based speech enhancement, they still require more than hundreds of MMACS. Given the limited computation budget of wearable devices like smart glasses, which is assumed to be around 100 MMACS, these approaches may not be feasible for deployment.

% efficient speech enhancement models
% Neural multi-channel speech enhancement models are usually designed to be quite large and computationally inefficient. Zoneformer~\cite{zoneformer} claims
% their models to be efficient but their computation still takes a few GMACS. Deepfilternet~\cite{mc_deepfilternet} based multi-channel speech enhancement model~\cite{mc_deepfilternet} is relatively efficient with 350 MMACS, enabling deployment on a RaspberryPi4~\cite{schröter2022deepfilternet2}. LLRNN~\cite{llrnn} explores time-domain small models as small hundreds of MMACS. However, for wearable devices like smart glasses, a desired computation budget would be around 50 MMACS or even less for possible deployment. 

% wearable devices and spear
%More recently, augmented hearing on smart glasses has become a popular topic~\cite{donley2021easycom, subspace_icassp, SPEAR, spatialanc}. Under noisy acoustic scenarios, conversations are undermined by environmental background noise and interference speakers. Augmented hearing aims to use a few microphones on smart glasses or headsets, to enhance a target conversation while reducing noise and inferences. SPEAR challenge~\cite{SPEAR} is proposed to facilitate the research on real-time speech enhancement with smart glasses, using the real-world EASYCOM dataset~\cite{donley2021easycom}. However, this challenge assumes ground-truth DoAs of all speakers are known.

Recently, there has been growing interest in augmented hearing on smart glasses \cite{donley2021easycom, subspace_icassp, SPEAR, spatialanc}. In noisy acoustic scenarios, conversations can be degraded by environmental background noise and interfering speakers. The objective of augmented hearing is to utilize a few microphones on smart glasses or headsets to enhance a target conversation while suppressing noise and interference. The SPEAR challenge \cite{SPEAR} was proposed to promote research on real-time speech enhancement with smart glasses using the real-world EASYCOM dataset \cite{donley2021easycom}. However, this challenge assumes that the ground-truth directions of arrival (DoAs) of all speakers are known, which is typically not the case in reality.  Instead, we propose a method for enhancing speech within a configurable conversational field of view (FoV) of a smart-glasses user, without the need for specific DoA information. This approach allows for more practical and realistic implementation of augmented hearing technology in real-world scenarios.

% In this work, we aim to address the following problems: (1) one single model that enhances all speakers with low distortion inside a configurable 2D Field of View (FoV), while reducing all sound sources outside of the FoV. (2) an ultra-low computation model with about 50 MMACS that suits power-constrained devices like smart glass. We propose a hybrid approach that combines neural networks, fixed beamforming, and adaptive beamforming. The configurable FoV enhancement method is crucial for on-glasses speech enhancement in daily speech communications, where a conversation usually happens in a horizontal field of view (FoV). This empowers smart glasses users to directly configure a speech enhancement with an FoV desired for an ongoing conversation. This also enables an automatic FoV speech enhancement controlled by other multi-modal scene analysis modules using, e.g., egocentric videos~\cite{donley2021easycom}. 

In this work, we aim to address two key problems in speech enhancement for smart glasses: (1) developing a single model that can enhance all speakers with low distortion within a configurable 2D field of view (FoV), while reducing all sound sources outside of the FoV; and (2) creating an ultra-low computation model with approximately 50 MMACS that is suitable for power-constrained devices like smart glasses. To achieve these goals, we propose a hybrid approach that combines neural networks, fixed beamforming, and adaptive beamforming. The proposed method allows for configurable FoV enhancement, which is crucial for on-glasses speech enhancement in daily speech communications where conversations typically occur within a horizontal field of view (FoV). This enables smart glasses users to directly configure speech enhancement for an ongoing conversation within a desired FoV, and also allows for automatic FoV speech enhancement controlled by other multi-modal scene analysis modules using egocentric videos \cite{donley2021easycom}.

\end{spacing}
\begin{figure}[t!]
\vspace{-1.5cm}
% \vspace{-1.5cm}
    \centering
    \includegraphics[width=0.85\linewidth]{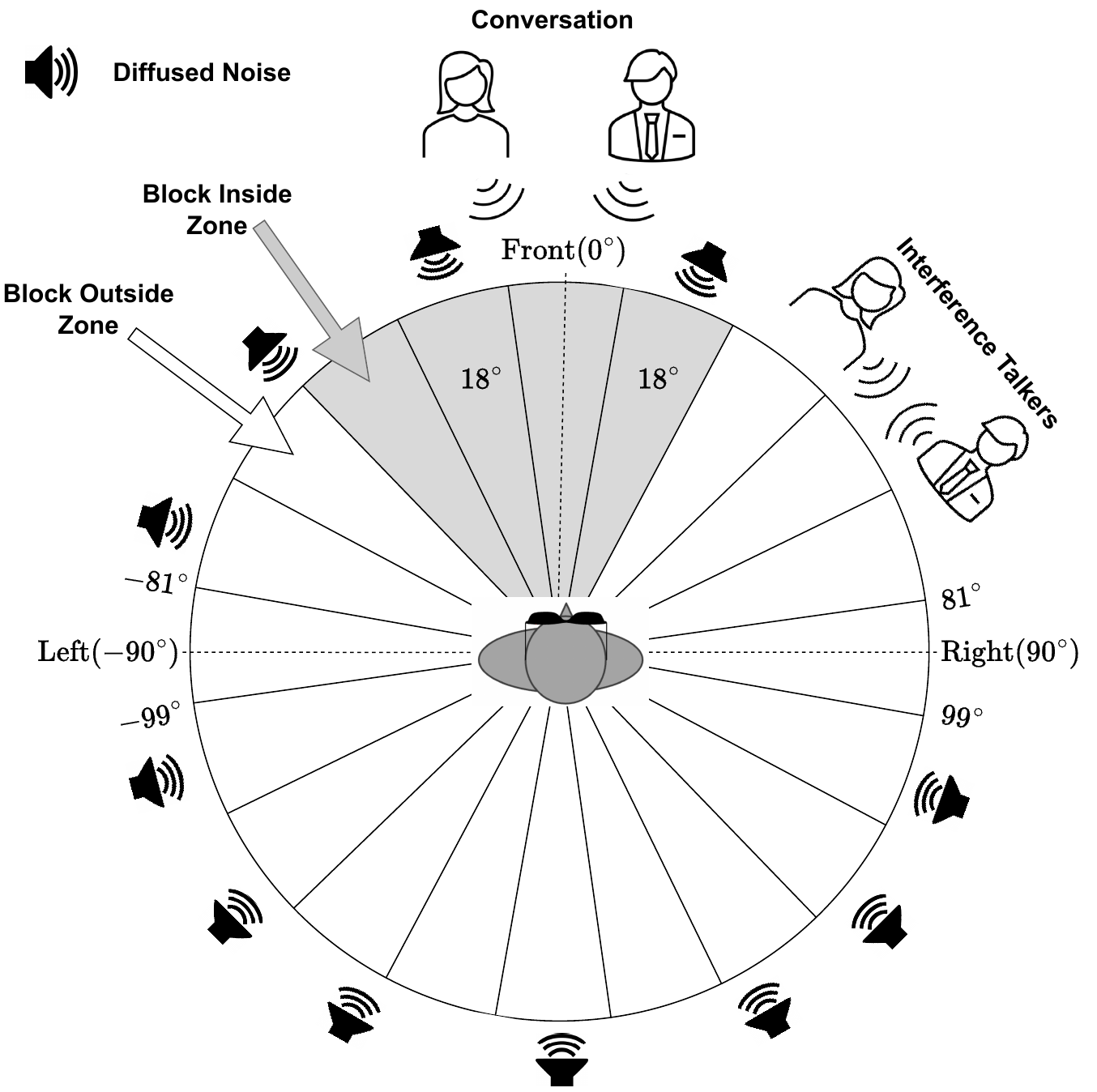}
    \vspace{-12pt}
    \caption{A user wearing a smart glasses with a mic-array. The horizontal plane is divided into $K=20$ blocks. The FoV (grey blocks) here is $-45^\circ$ to $27^\circ$, containing the target conversation.}
    \label{fig:fov}
    \vspace{-17pt}
\end{figure}

\vspace{-8pt}
\section{Problem Formulation}
\vspace{-5pt}
\label{sec:problem}
\begin{spacing}{0.95}
We considered a microphone array mounted on smart glasses, as shown in Figure~\ref{fig:fov}. We assume the smart glasses are equipped with $M$ microphones, and they mostly lie on a horizontal plane. We divide the $360^\circ$ horizontal plane around a user into $K$ discretized spatial blocks each containing a field of view (FoV) of $(\frac{360}{K})^\circ$, exactly as shown in Figure~\ref{fig:fov} with $K=20$. The target conversation could happen in some FoV represented by a few consecutive blocks, e.g., the grey area in Figure~\ref{fig:fov}. These consecutive spatial blocks can be directly configured by a user, or controlled by other multi-modal scene analysis modules. Note that the FoV here is defined to be in the format of consecutive spatial blocks and therefore could cover an arbitrary continuous region, for example, the region could be configured to enhance talkers in front of the user, e.g., -45 to 45, talkers sitting beside the user, e.g., -90 to -45, or a combination of such cases, e.g., -90 to 90. Our model's inputs are (1) the set of block indices $I_{\text{FoV}} \triangleq \{[i, j], 1 \leq i \leq j \leq K, i, j\in\mathbb{Z}\}$ that represents a configurable target FoV, and (2) the $M$-channel recorded noisy audio's STFT $X\in\mathbb{R}^{M\times T \times F}$, where $T$ represents number of frames and $F$ represents number of frequency bins. Our model's output is defined to be a single-channel stream that contains all speech signals inside the configured FoV.

Our model consists of the following components: (1) feature extraction consisting of spatial feature and ref-channel feature (2) FoVNet which is an FoV conditioned network that enhances target speech signals in the FoV by estimating an ERB gain, which is psychoacousticly desired for human hearing (3) low-distortion multi-channel Wiener filter and post-processing. The overall enhancement pipeline is shown in Figure~\ref{fig:pipeline}.

\end{spacing}
\vspace{-8pt}
\section{Method}
% \vspace{-5pt}
\begin{spacing}{0.95}
% \vspace{-5pt}
% \begin{figure}[t!]
% \vspace{-0.6cm}
%     \centering
%     \includegraphics[width=0.9\linewidth]{figures/fov_pipeline.pdf}
%     \vspace{-8pt}
%     \caption{Configurable FoV Enhancement Pipeline.} 
%     \label{fig:pipeline}
%     \vspace{-15pt}
% \end{figure}
\vspace{-5pt}
\subsection{Feature Extraction}
\vspace{-4pt}
\label{sec:feature}
{\bf Spatial Feature}: maxDI beamformer~\cite{donley2021easycom} is a fixed MVDR beamformer assuming an isotropic diffused noise field. Previous studies have shown the effectiveness of it as a front-end feature extractor for multi-channel speech enhancement~\cite{Benjamin, tabe, beamspace}. We also adopt it as a spatial feature extractor to spatially sample incoming signals around the $360^\circ$ horizontal plane. Thus we sample each spatial block defined in Section~\ref{sec:problem} with a maxDI beamformer. Assume the $k^{th}$ spatial block's center angle is $\theta_{k}$, then a fixed maxDI beamformer $w_{\theta_k}(f)\in\mathbb{C}^{M\times{1}}$ is designed to beamform towards $\theta_k$. Then the $k^{th}$ block's feature $s_k\in\mathbb{R}^{T\times 64}$ would be the log scale 64-ERB band of the corresponding beamformer result:
% \begin{align}
%     b_k(t, f) &= w_{\theta_k}(f)^HX(t, f)\\
%     s_k &= \text{log}(\text{ERB}_{64}(b_k))
% \end{align}
\vspace{-4pt}
\begin{equation}
    b_k(t, f) = w_{\theta_k}(f)^H X(t, f); s_k = \log(\text{ERB}_{64}(b_k))
    \vspace{-4pt}
\end{equation}
where $X(t, f)\in\mathbb{C}^{M\times 1}$ is the noisy multi-channel signal's STFT, $b_k\in\mathbb{C}^{T\times F}$ is the $k^{th}$ beamformer's STFT output, and $\text{ERB}_{64}$ is the 64-ERB filterbank transform to get the final spatial feature $s_k\in\mathbb{R}^{T\times 64}$ for the $k^{th}$ block. By concatenating all $K$ blocks' features $\{s_1, s_2, ..., s_K\}$, we get the final spatial feature $S\in\mathbb{R}^{K\times T\times 64}$, and $S(k, t, b)$ represents the $b^{th}$ ERB band at $k^{th}$ block and $t^{th}$ frame.
% Besides the spatial features, we also extract the reference channel's 64-band log ERB feature. 

{\bf Reference-channel Feature}: Because later the neural network estimates a denoising band gain that applies on the reference channel, the reference channel's noisy audio feature is also extracted as input of the neural network. Assume the reference channel's noisy signal STFT is $X_{\text{ref}}(t,f)$, then the reference-channel feature $R = \text{log}(\text{ERB}_{64}(X_{\text{ref}})\in\mathbb{R}^{T\times 64}$, where $R(t, b)$ represents the $b^{th}$ ERB band feature at frame $t$.

When calculating STFT, we use a hop size of 128, a FFT size of 256, a frame size of 256, and a hanning window.
\begin{figure}[t!]
% \vspace{-1.5cm}
\vspace{-1.5cm}
    \centering
    \includegraphics[width=0.85\linewidth]{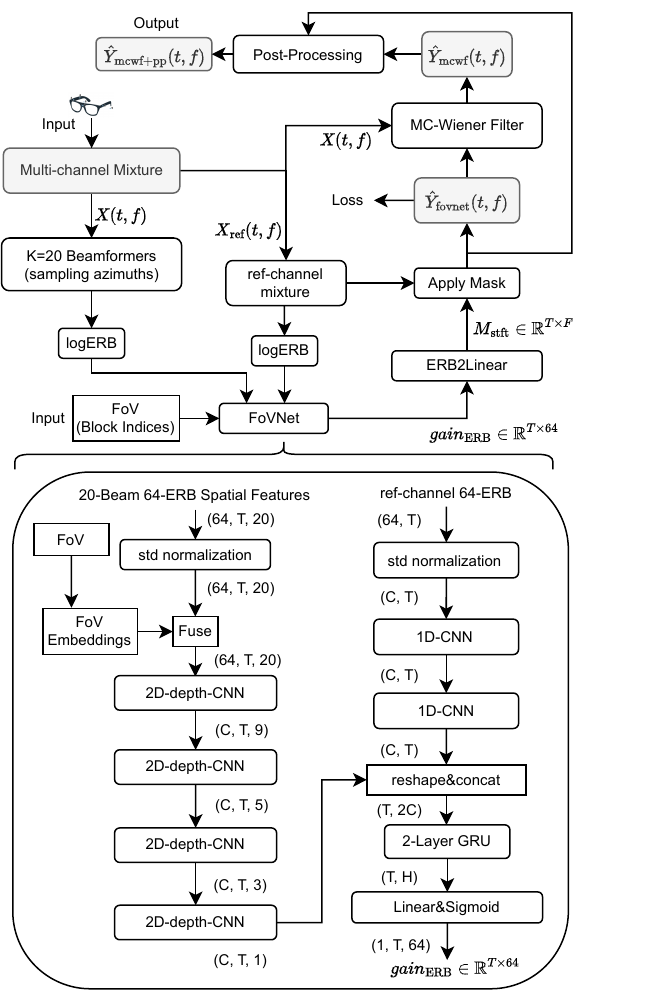}
    \vspace{-8pt}
    \caption{Configurable FoV Enhancement Pipeline.} 
    \label{fig:pipeline}
    \vspace{-20pt}
\end{figure}
\vspace{-8pt}
\subsection{FoVNet}
\vspace{-5pt}
{\bf Inputs and Normalization}: As shown in Figure~\ref{fig:pipeline}, the neural network takes the spatial features $S\in\mathbb{R}^{K\times T\times 64}$, ref-channel feature $R\in\mathbb{R}^{T\times 64}$, and the FoV as inputs. In our setting, we set $K=20$. The spatial features are first normalized with mean zero and unit variance with pre-calculated statistics. The same is done for ref-channel features. We denote the normalized spatial feature and ref-channel feature as $S_{\text{norm}}$ and $R_{\text{norm}}$, respectively. The FoV input is represented as a set of block indices $I_{\text{FoV}} $, where any element in this set corresponds to a spatial block that's inside the FoV, as mentioned in Section~\ref{sec:problem}.

{\bf FoV Embedding}:
The normalized spatial feature $S_{\text{norm}}$ contains features of $K=20$ spatial blocks, as explained in Section~\ref{sec:feature}. From $K$ spatial blocks, $I_{\text{FoV}}$ contains indices of the spatial blocks considered inside FoV. To encode the FoV information inside the neural network, we design two sets of learnable embeddings indicating whether a block is inside FoV or outside of FoV. The two sets of embeddings are $\{E^{\text{in}}_{\mu}, E^{\text{in}}_{\sigma}\in\mathbb{R}^{64}\}$ and $\{E^{\text{out}}_{\mu}, E^{\text{out}}_{\sigma}\in\mathbb{R}^{64}\}$. Then similar to FilM~\cite{perez2017film}, the embeddings are fused inside the normalized spatial feature:
% \begin{align}
%     % S_{\text{norm}}(k, t, :) \longleftarrow k \in I_{\text{FoV}}\\
%     % \[
% S_{\text{norm}}(k, t, :) \leftarrow
% \begin{cases} 
% S_{\text{norm}}(k, t, :) \odot E^{\text{in}}_{\sigma} \oplus E^{\text{in}}_{\mu}& \text{if } k \in I_{\text{FoV}}\\
% S_{\text{norm}}(k, t, :) \odot E^{\text{out}}_{\sigma} \oplus E^{\text{out}}_{\mu} & \text{if } k \notin I_{\text{FoV}}.
% \end{cases}
% \end{align}
\vspace{-8pt}
\begin{align}
S_{\text{norm}}(k, t, :) \leftarrow
\begin{cases} 
S_{\text{norm}}(k, t, :) \odot E^{\text{in}}_{\sigma} \oplus E^{\text{in}}_{\mu}& k \in I_{\text{FoV}}\\
S_{\text{norm}}(k, t, :) \odot E^{\text{out}}_{\sigma} \oplus E^{\text{out}}_{\mu} & k \notin I_{\text{FoV}}
\vspace{-2pt}
\end{cases}
\vspace{-5pt}
\end{align}
where $\odot$ and $\oplus$ corresponds to element-wise multiplication and addition. Note that all the embeddings are learnable.

{\bf FoVNet Architecture}
\label{sec:architecture}
To process the normalized and FoV-fused spatial features, four 2-D depth-wise convolutional layers are applied to the spatial features. These convolutions are performed across both the T-dimensional time domain and the K-dimensional spatial domain, considering the dimension of the ERB band (64) as the channel dimension. The convolutional kernels are configured as (2, 3) for temporal and spatial dimensions, and similarly, the strides are set to be (1, 2). The output channel dimension of each layer is set to $C=80$. BatchNorm~\cite{ioffe2015batch} and leaky ReLU~\cite{lrelu} (0.1 negative slope) are used as normalization layers and activations. The four convolutional layers keep the time dimension uncompressed, but keep compressing the spatial dimension. In our case, $K=20$, with proper padding, the final spatial dimension becomes 1 after four layers, as shown in Figure~\ref{fig:pipeline}.

To process the normalized, ref-channel feature, two 1-D convolutional layers are applied. The 1-D convolution is applied to the temporal dimension with kernel size 3. Again, the 64-dimensional ERB features are treated as the channel. The output channel dimension of both layers is all set to be $C=80$.

After the CNN encodings, the spatial CNN branch's output and the ref-channel CNN branch's output are concatenated in the channel dimension. The concatenated feature has a dimension $T\times2C$, which is then sequentially processed by a 2-layer GRU with hidden dimension $H=96$. Then a single $96\times64$ linear layer with sigmoid activation transforms each time step's $H$-dimensional feature into the final $64$-dimensional ERB gain. We denote the final ERB gain as $\text{gain}_{\text{ERB}} \in \mathbb{R}^{T \times 64}$. 
$\text{gain}_{\text{ERB}}$ is then transformed back to the linear frequency scale as an STFT magnitude mask $M_{\text{stft}} \in \mathbb{R}^{T \times F}$ which is applied to the reference channel noisy STFT $X_{\text{ref}}(t, f)$ to get the estimated clean speech STFT $\hat{Y}_{\text{fovnet}}(t, f)$. Then after inverse STFT (ISTFT), we recover the estimated time-domain speech $\hat{y}(t)$.

{\bf Training} 
We use SI-SDR~\cite{roux2018sdr} loss and an STFT loss:
% \vspace{-1pt}
% \begin{align}
% \label{eq:loss}
% \footnotesize
%  \text{loss} &= -\text{SI-SDR}(y, \hat{y}) + 
%  \lambda_1 \left\Vert \log\left(\left| \text{stft}(y) \right|\right) - \log\left(\left| \text{stft}(\hat{y}) \right|\right) \right\Vert_1\nonumber\\
%  &+ \lambda_2
%     \vspace{-1pt}
% \end{align}
{\footnotesize\begin{align}
\label{eq:loss}
 \text{loss} &= -\text{SI-SDR}(y, \hat{y}) + 
 \lambda_1 \left\Vert \log\left(\left| Y \right|\right) - \log\left(\left| \hat{Y}_{\text{fovnet}} \right|\right) \right\Vert_1\nonumber\\
 &+ \lambda_2 \left\Vert \log\left(\left| \text{Re}(Y) \right|\right) - \log\left(\left| \text{Re}(\hat{Y}_{\text{fovnet}}) \right|\right) \right\Vert_1\nonumber\\
 &+ \lambda_2 \left\Vert \log\left(\left| \text{Im}(Y) \right|\right) - \log\left(\left| \text{Im}(\hat{Y}_{\text{fovnet}}) \right|\right) \right\Vert_1
    \vspace{-1pt}
        \end{align}}
where $y$ is assumed as the time-domain target speech signal (mixture containing all the speech signals inside FoV) and $\hat{y}$ as the estimation. $Y$ and $\hat{Y}_{\text{fovnet}}$ are the short time fourier transform (STFT) of $y$ and $\hat{y}$ respectively. Re and Im denote real and imaginary parts. For STFT, we use a hop size of 128, an FFT size of 256, a frame size of 256, and a hanning window. We also set up $\lambda_1=0.01, \lambda_2=1$ in Eq.\ref{eq:loss}.

This network is computationally efficient, i.e., has a computation footprint of 50 MMACS. It uses the ERB band features to compress the spectral dimension and uses a novel FoV embedding to encode FoV information in a computationally light way. However, the network only enhances the magnitude of the speech signal. Since the network is tiny, has a small computation capacity, and only enhances magnitude, the network output is likely distorted in low-snr acoustic conditions. Thus we propose to further use a multi-channel Wiener filter to improve enhancement performance while controlling speech distortion.

\vspace{-8pt}
\subsection{Multi-channel Wiener Filter and Post-processing}
\vspace{-5pt}
{\bf Multi-channel Wiener Filter}: we further aim at improving noise reduction performance while controlling speech distortion by using a multi-channel Wiener filter with low-distortion beamforming~\cite{dnnwienerdnn1, dnnwienerdnn2}. So far we have the neural network enhanced ref-channel signal STFT $\hat{Y}_{\text{fovnet}}(t, f)\in\mathbb{C}^{1\times1}$, along with the original multi-channel mixture STFT $X(t, f)\in\mathbb{C}^{M\times1}$. Thus we can estimate a smoothed noisy signal covariance $\hat{\Phi}_{xx}(t, f)$ and a smoothed cross-covariance $\hat{\Phi}_{xy}(t, f)$ by:
\vspace{-7pt}
{\scriptsize\begin{align}
\hat{\Phi}_{xx}(t, f) &= (1-\alpha_{xx})\hat{\Phi}_{xx}(t-1, f) + \alpha_{xx} X(t, f)X^H(t, f)\\
% Equation 12
\hat{\Phi}_{xy}(t, f) &= (1-\alpha_{xy})\hat{\Phi}_{xy}(t-1, f) + \alpha_{xy} X(t, f)\hat{Y}_{\text{fovnet}}(t, f)
\\[-15pt]\nonumber
\end{align}}

$\alpha_{xx}$, and $\alpha_{xy}$ are recursive update coefficients. We empirically set $\alpha_{xx}=0.01$, and $\alpha_{xy}=0.03$. Thanks to smoothed covariance matrices, we derive a low-distortion multi-channel Wiener filter $h_{\text{mcwf}}(t, f)\in\mathbb{C}^{M\times1}$ solution as $\hat{y}_{\text{mcwf}}$:
\vspace{-5pt}
\begin{align}
    h^H_{\text{mcwf}}(t, f) &= \hat{\Phi}^{-1}_{xx}(t, f)\hat{\Phi}_{xy}(t, f)\\
    \hat{Y}_{\text{mcwf}}(t, f) &= h^H_{\text{mcwf}}(t, f)X(t, f)
    \\[-20pt]\nonumber
\end{align}

{\bf Post Processing}: Although the multi-channel Wiener filtering would reduce speech distortion, it also preserves more residual noise. We thus reuse $\hat{Y}_\text{fovnet}(t, f)$ to do the post-processing:
\vspace{-5pt}
\begin{align}
    \label{eq: newmask}
    \hat{M}_{\text{stft}}(t, f) &= \max\left(\min\left(1, \frac{|\hat{Y}_\text{fovnet}(t, f)|}{|\hat{Y}_{\text{mcwf}}(t, f)|}\right), \epsilon\right)\\
    \label{eq: mcwf+pp}
    \hat{Y}_{\text{mcwf+pp}} &= \hat{M}_{\text{stft}}(t, f) \cdot \hat{Y}_{\text{mcwf}}(t, f)
    \\[-20pt]\nonumber
\end{align}
Eq.~\ref{eq: newmask} creates a new STFT mask $\hat{M}_{\text{stft}}(t, f)$ for post-processing. $\epsilon$ denotes a small floor value to avoid aggressive denoising. We set $\epsilon$ to be $0.1$. The minimum operation acts like a selector between $|\hat{Y}_{\text{mcwf}}|$, and $|\hat{Y}_{\text{fovnet}}|$. Then after masking, the magnitude of the final output $\hat{Y}_{\text{mcwf+pp}}$ should be the minimum between $|\hat{Y}_{\text{fovnet}}(t, f)|$, and $|\hat{Y}_{\text{mcwf}}(t, f)|$, if $\epsilon$ is $0$, which aims to remove residual noise in $|\hat{Y}_{\text{mcwf}}|$.

\begin{table*}[ht]
\vspace{-1.35cm}
\centering
% \vspace{-20pt}
\caption{Evaluation results for single target talker scenarios. (Top 3 methods bolded)}
\vspace{-5pt}
\scriptsize
\begin{tabular}{|c|c|c|c|c|c|c|c|c|c|c|c|c|}
\hline
\textbf{Models/Metrics} & \multicolumn{3}{l|}{\textbf{1 Target 0 Interference}} & \multicolumn{3}{l|}{\textbf{1 Target 1 Interference}} & \multicolumn{3}{l|}{\textbf{1 Target 2 Interference}} & \multicolumn{3}{l|}{\textbf{1 Target 3 Interference}} \\ \hline
                        & PESQ & STOI & SI-SDR & PESQ & STOI & SI-SDR & PESQ & STOI & SI-SDR & PESQ & STOI & SI-SDR \\ \hline
Noisy                   & 1.46 & 0.64 & -1.25 & 1.42 & 0.59 & -3.08 & 1.36 & 0.56 & -3.91 & 1.31 & 0.53 & -5.16 \\ \hline
SC-CRN                  & 1.78 & 0.67 & 3.41  & 1.57 & 0.60 & -0.98 & 1.47 & 0.56 & -2.12 & 1.38 & 0.52 & -3.70 \\ \hline
MC-CRN                  & 1.92 & 0.72 & 4.88  & 1.62 & 0.64 & 0.21  & 1.53 & 0.61 & -0.75 & 1.44 & 0.56 & -2.55 \\ \hline
maxDI                & 1.71 & 0.72 & 0.60  & 1.66 & 0.69 & -0.34 & 1.59 & 0.67 & -1.03 & 1.54 & 0.64 & -2.24 \\ \hline
maxDI + SC-CRN       & \textbf{2.06} & \textbf{0.76} & \textbf{4.21}  & \textbf{1.90} & \textbf{0.72} & \textbf{2.34}  & \textbf{1.81} & \textbf{0.69} & {1.63}  & \textbf{1.75} & \textbf{0.66} & 0.33  \\ \hline
FoVNet                 & \textbf{2.02} & \textbf{0.74} & \textbf{5.53}  & \textbf{1.93} & \textbf{0.72} & \textbf{4.04}  & \textbf{1.86} & \textbf{0.70} & \textbf{3.50}  & \textbf{1.77} & \textbf{0.67} & \textbf{2.33}  \\ \hline
FoVNet + MCWF          & 1.91 & 0.71 & 3.55  & 1.85 & 0.69 & 2.39  & 1.79 & 0.67 & \textbf{1.85}  & 1.72 & 0.63 & \textbf{0.77}  \\ \hline
FoVNet + MCWF + PP     & \textbf{2.05} & \textbf{0.74} & \textbf{5.01}  & \textbf{1.99} & \textbf{0.72} & \textbf{3.80}  & \textbf{1.92} & \textbf{0.70} & \textbf{3.33}  & \textbf{1.85} & \textbf{0.67} & \textbf{2.23}  \\ \hline
\end{tabular}
\label{tab:result1}

\vspace{5pt}
\caption{Evaluation results for double target talker scenarios. (Top 3 methods bolded)}
\vspace{-5pt}
\scriptsize
\begin{tabular}{|c|c|c|c|c|c|c|c|c|c|c|c|c|}
\hline
\textbf{Models/Metrics} & \multicolumn{3}{l|}{\textbf{2 Target 0 Interference}} & \multicolumn{3}{l|}{\textbf{2 Target 1 Interference}} & \multicolumn{3}{l|}{\textbf{2 Target 2 Interference}} & \multicolumn{3}{l|}{\textbf{2 Target 3 Interference}} \\ \hline
                        & PESQ & STOI & SI-SDR & PESQ & STOI & SI-SDR & PESQ & STOI & SI-SDR & PESQ & STOI & SI-SDR \\ \hline
Noisy                   & 1.78 & 0.75 & 4.31  & 1.56 & 0.66 & 0.64  & 1.53 & 0.64 & 0.14  & 1.45 & 0.60 & -1.01 \\ \hline
SC-CRN                  & 2.15 & 0.77 & 6.80  & 1.78 & 0.66 & 2.21  & 1.70 & 0.64 & 1.46  & 1.58 & 0.59 & 0.18  \\ \hline
MC-CRN                  & 2.27 & 0.81 & 7.66  & 1.84 & 0.70 & 3.17  & 1.77 & 0.68 & 2.53  & 1.63 & 0.63 & 1.02  \\ \hline
maxDI                & 1.96 & 0.80 & 4.59  & 1.79 & 0.73 & 2.24  & 1.77 & 0.73 & 2.27  & 1.67 & 0.69 & 1.16  \\ \hline
maxDI + SC-CRN            & \textbf{2.34} & \textbf{0.82} & \textbf{6.25}  & \textbf{2.07} & \textbf{0.74} & \textbf{3.93}  & \textbf{2.02} & \textbf{0.74} & \textbf{3.86}  & \textbf{1.89} & \textbf{0.70} & 2.76  \\ \hline
FoVNet            & \textbf{2.36} & \textbf{0.83} & \textbf{8.05}  & \textbf{2.10} & \textbf{0.76} & \textbf{5.33}  & \textbf{2.08} & \textbf{0.76} & \textbf{5.22}  & \textbf{1.98} & \textbf{0.73} & \textbf{4.26}  \\ \hline
FoVNet + MCWF                 & 2.11 & 0.80 & 6.06  & 1.94 & 0.73 & 3.87  & 1.92 & 0.73 & 3.81  & 1.84 & 0.69 & \textbf{2.83}  \\ \hline
FoVNet + MCWF + PP            & \textbf{2.34} & \textbf{0.82} & \textbf{7.32}  & \textbf{2.13} & \textbf{0.76} & \textbf{4.96}  & \textbf{2.11} & \textbf{0.76} & \textbf{5.02}  & \textbf{2.01} & \textbf{0.73} & \textbf{4.06}  \\ \hline
\end{tabular}
\label{tab:result2}
\vspace{-15pt}
\end{table*}

\begin{table}[h]
\vspace{-1pt}
\centering
\caption{Compare models in MMACs and number of parameters}
\vspace{-5pt}
\scriptsize 
\begin{tabular}{|c|c|c|c|c|}
\hline
\textbf{Models} & \textbf{MMACS} & \textbf{Params(M)} & \textbf{Config} & \textbf{DoA} \\ \hline
SC-CRN                  & 49.12          & 0.183           & $\times$                    & $\times$               \\ \hline
MC-CRN                  & 48.06          & 0.165           & $\times$                     & $\times$               \\ \hline
maxDI                  & -          & -           & $\times$                     & $\checkmark$               \\ \hline
maxDI + SC-CRN                  & 49.12          & 0.183           & $\times$                     & $\checkmark$               \\ \hline
FoVNet                 & 49.09          & 0.206           & $\checkmark$                     & $\times$               \\ \hline
FoVNet + MCWF                  & 49.09          & 0.206           & $\checkmark$                     & $\times$               \\ \hline
FoVNet + MCWF + PP                  & 49.09          & 0.206           & $\checkmark$                     & $\times$               \\ \hline
% Add the rest of the models here
\end{tabular}
\label{tab:configs}
\vspace{-21pt}
\end{table}

\end{spacing}
\vspace{-8pt}
\section{Experiments and Results}
\vspace{-5pt}
\begin{spacing}{0.95}
\subsection{Dataset}
\vspace{-5pt}
\label{sec:dataset}
We use clean speech and noise datasets from the first DNS challenge~\cite{dns1}. We considered the 5-channel microphone array mounted on prototype smart glasses, similar to the one in EasyCom~\cite{donley2021easycom}. All audio samples are synthesized at a 16kHz sampling rate. For room acoustics and array setup, we simulate 40,000 "shoebox" rooms with dimensions ranging from 3x3x3 to 10x10x4 meters, alongside 10 distinct array positions and orientations per room, using Pyroomacoustics~\cite{pyroomacoustics}. For each sample in the dataset, the following steps are taken: {(1) \bf FoV Sampling}: The size of FoV is randomly sampled from 2 to 10 blocks (18 degrees each), then based on the sampled block size, the specific spatial blocks are randomly sampled but with one constraint that none of the blocks are $\leq-99^\circ$ or $\geq99^\circ$. The setup aligns with Figure~\ref{fig:fov}, where we assume the FoV should be in the front semi-circle. {(2) \bf Signal Sampling}: randomly 1 to 50 noise sources, {\bf 0 to 3 interference talkers}, and { \bf 1 to 2 target talkers} are sampled per dataset sample. Noise samples are randomly placed anywhere. All speakers are placed at -30 to 30 degrees in elevation. Interference speakers are randomly placed outside the field of view (FoV), and maintain a minimum of 10 degrees in azimuth separation from the FoV. Target talkers are randomly placed within the FoV. {(3) \bf Mixture synthesis}: We synthesize the noisy sample with Pyroomacoustics using the sampled room, array position, array orientation, noise/speech signals, and corresponding signal positions. We also set the signal-to-noise ratio (SNR) to be randomly from $-10$ dB to $5$ dB, and the signal-to-interference ratio (SIR) to be between $-2$ dB and $2$ dB. Overall we generate 80,000 10-second samples for training, 3,000 for validation, and 3,000 for testing.

% Acoustic Parameters: Samples offer a signal-to-noise ratio (SNR) from -10 dB to 5 dB and, when applicable, a signal-to-interference ratio (SIR) between -2 dB and 2 dB. Duration for each audio sample is set at 10 seconds.

% Acoustic Parameters: Samples offer a signal-to-noise ratio (SNR) from -10 dB to 5 dB and, when applicable, a signal-to-interference ratio (SIR) between -2 dB and 2 dB. Duration for each audio sample is set at 10 seconds.

% Environmental Variability: The dataset simulates 40,000 "shoebox" rooms with dimensions ranging from 3x3x3 to 10x10x4 meters, alongside 10 distinct array positions and orientations per room.

% {(1) \bf Mixture synthesis}:: Using Pyroomacoustics~\cite{} with the image source method, we generate 80,000 training, 3,000 validation, and 3,000 testing samples. The synthesis involves detailed steps from FoV selection to room and array positioning, integrating speech and noise sources according to outlined constraints for realistic acoustic simulations.

% Designed for advanced audio signal processing algorithm development and evaluation, this dataset provides a broad array of synthesized acoustic scenarios to emulate real-world audio environments accurately.

\vspace{-10pt}
\subsection{Models and Baselines}
\vspace{-5pt}
All the baselines and proposed methods are shown in Table~\ref{tab:configs}, with all configurations. \textbf{MMACS} represents model complexity in million multiply-add per second, \textbf{Params} represents the number of million parameters, \textbf{Config} represents whether the method is FoV configurable, and \textbf{DoA} represents whether the method needs targets' DoA information as input.

The baseline models include the single-channel Convolutional Recurrent Network (\textbf{SC-CRN}) and multi-channel CRN (\textbf{MC-CRN}), both employing 2D depth-convolutions across time and ERB band dimension to estimate ERB band gains. The architectures are very similar to CRN~\cite{crn} except that we substitute 2D-CNN with 2D depth-wise CNN and compress the model size to be about 50 MMACS. The \textbf{SC-CRN} only uses the ref-channel feature as input. The \textbf{MC-CRN} utilizes the concatenated spatial and ref-channel features but treats the $K+1$-dimensional ($K$-spatial, $1$-ref-channel) concatenated dimension as the channel dimension. They are trained with the same objective function, but to enhance all speech components (both target and interference). Thus SC-CRN and MC-CRN are not able to distinguish targets and interferences.

% Additionally, for the case with a single target talker, we use \textbf{ISO-MVDR}~\cite{SPEAR} as a baseline, assuming the target DoA is given. For the case with two target talkers, we use \textbf{maxDI} beamforming as a baseline, which is an extension of ISO-MVDR except now we want to enhance two target talkers. maxDI sets distortionless constraints on two target talkers' DoAs, which would enhance both speakers. The ISO-MVDR and maxDI results are further denoised using SC-CRN to have two other baselines \textbf{ISO-MVDR + SC-CRN} and \textbf{maxDI + SC-CRN}. Note that, unlike our methods, ISO-MVDR and maxDI based baselines all assume known target direction(s), and thus can distinguish target and interferences and adopt most information.
Additionally, we use \textbf{maxDI} beamformer as a baseline, assuming target direction(s) are known. For the case of a single target talker, the maxDI beamformer is just the MVDR beamformer with isotropic diffused noise. For the case of two target talkers, we define the maxDI beamformer to be the LCMV beamformer with two known DoAs' distortionless constraints~\cite{beamforming} under isotropic diffused noise. maxDI beamformer's output is further denoised using SC-CRN to have another baseline \textbf{maxDI + SC-CRN}. Note that, unlike our proposed methods, maxDI-based baselines all are aware of ground-truth DoAs of target talkers, and thus can distinguish targets and interferences, adopting the most spatial information.

Our proposed models include the \textbf{FoVNet}'s result $\hat{Y}_{\text{fovnet}}$, \textbf{FoVNet + MCWF}'s result $\hat{Y}_{\text{mcwf}}$, and \textbf{FoVNet + MCWF + PP}'s result $\hat{Y}_{\text{mcwf+pp}}$. All the networks are trained with ADAM optimizer~\cite{kingma2017adam} with 200 epochs with a learning rate of $2\times10^{-4}$. Also, all neural models' convolutional layers have causal padding such that all methods only have 16 ms of latency.
\end{spacing}
\vspace{-9pt}
\section{Results and Discussions}
\vspace{-6pt}
\begin{spacing}{0.95}
We evaluate our methods and baselines on our test set as described in Section~\ref{sec:dataset}, with 1-2 target talkers and 0-3 interference speakers. We use SI-SDR, PESQ(NB), and STOI as evaluation metrics.Table~\ref{tab:result1} and Table~\ref{tab:result2} show the results of a single target talker case and the two target talkers case, respectively.

{\bf FoVNet VS. SC-CRN and MC-CRN}: Since SC-CRN and MC-CRN are not able to distinguish target and interference speakers, we compare them with our methods in the setting of $0$ interference speaker scenario. We observe that in both single and double target talker(s) scenarios, our FoVNet performs significantly better than MC-CRN and SC-CRN by a large margin in all evaluation metrics, indicating how FoVNet succeeds in exploiting the configurable FoV information.

{\bf FoVNet VS. maxDI-based methods}: Although maxDI beamformer benefits from the ground-truth DoAs of target talkers, FoVNet performs better overall, in particular for non-zero interference speakers. For the case of a single target talker, FoVNet performs much better than maxDI in all metrics because maxDI has no neural network processing. With interference speakers, FoVNet performs slightly better than maxDI + SC-CRN, which benefits from ground-truth DoAs. Without interference speakers, FoVNet is 1.3dB better than maxDI+SC-CRN in SI-SDR, but 0.04 and 0.02 slightly worse than maxDI+SC-CRN in terms of PESQ and STOI, showing very similar results. For the case of two target talkers, FoVNet is better than maxDI+SC-CRN for all cases and metrics. One observation is that with more interference speakers, the performance gap between FoVNet and maxDI+SC-CRN becomes larger. This can be attributed to the failure of maxDI to adequately reduce interferences. The conclusion is that overall our FoVNet performs better than maxDI+SC-CRN even though FoVNet does not benefit from ground truth target DoA(s).

{\bf MCWF and Post-Processing}: We also compare FoVNet's performance with that of FoVNet+MCWF and FoVNet+MCWF+PP. Through observation, MCWF suffers in all metrics, which is attributed to residual noise/interference. However, we observe FoVNet+MCWF+PP has a consistent performance gain in PESQ compared with FoVNet in all cases except the case of two target talkers and zero interference. The gain is more obvious for a single target speaker case. From all metrics, PESQ is most correlated to speech perceptual quality, which shows the effectiveness of MCWF+PP in perceptual quality improvement.
\end{spacing}
\vspace{-9pt}
\section{Conclusion}
\vspace{-5pt}
\begin{spacing}{0.95}
% This work introduces a novel multi-channel approach for smart glasses, combining deep learning with signal processing for efficient, configurable FoV speech enhancement. Experimental results show the effectiveness of FoVNet in generalizing to flexible FoVs while maintaining superior enhancement performance. Excelling in computational efficiency and speech quality across multiple scenarios, it showcases the potential for augmented hearing on power-sensitive devices, representing a significant leap in wearable speech enhancement technology.
This work introduces a novel multi-channel approach for smart glasses that combines deep learning and signal processing techniques to achieve efficient, effective, and configurable speech enhancement within the field of view (FoV) of users for daily speech communication, without the need for specific DoA information. The proposed FoVNet demonstrates superior performance in enhancing speech while generalizing to various flexible FoVs. Experimental results showcase the potential of FoVNet for augmented hearing on power-sensitive devices, representing a significant advancement in wearable speech enhancement technology. FoVNet excels in both computational efficiency and speech quality across multiple scenarios, making it a promising solution for wearable smart glasses applications.

\end{spacing}

\bibliographystyle{IEEEtran}
\begin{spacing}{1}
\bibliography{mybib}
\end{spacing}

\end{document}